\documentclass[usenatbib,usegraphicx]{mn2e}
\usepackage{amsfonts}
\usepackage{amsmath}
\usepackage{amssymb}
\usepackage{comment}
\usepackage{color}

\title[A \it{Herschel}-ATLAS study of local minor mergers]
    {\vspace{-0.25in} A {{\it Herschel}\thanks{{\it Herschel} is an ESA space observatory with science instruments provided by European-led Principal Investigator consortia
and with important participation from NASA.}-ATLAS study of dusty
spheroids: probing the minor-merger process in the local Universe
\vspace{-0.25in}}}

\author[Sugata Kaviraj et al.]
{S. Kaviraj,\thanks{E-mail: skaviraj@astro.ox.ac.uk}$^{1,2,3}$ K.
Rowlands,$^{4}$ M. Alpaslan,$^{5,6}$ L. Dunne,$^{4,7}$ Y. S. Ting,$^{2}$
\newauthor M. Bureau,$^{2}$ S. Shabala,$^{8}$ C. J. Lintott,$^{2}$ D. J. B. Smith,$^{1}$ N. Agius,$^{9}$ R. Auld,$^{10}$ \newauthor M. Baes,$^{11}$ N.
Bourne,$^{4}$ A. Cava,$^{12}$ D. L. Clements,$^{3}$ A. Cooray,$^{13}$ A.
Dariush,$^{3}$ \newauthor G. De Zotti,$^{14,15}$ S. P. Driver,$^{5,6}$ S.
Eales,$^{10}$ R. Hopwood,$^{3}$ C. Hoyos,$^{4}$ E. Ibar,$^{16}$ \newauthor
S. Maddox,$^{4,7}$ M.~J.~Micha{\l}owski,\thanks{FWO Pegasus Marie
Curie Fellow}$^{11,6}$ A. E. Sansom,$^{9}$ M. Smith$^{10}$ and E. Valiante$^{10}$\\\\
$^{1}$Centre for Astrophysics, Science \& Technology Research Institute,
University of Hertfordshire, Hatfield, Herts, AL10 9AB, UK\\
$^{2}$Department of Physics, University of Oxford, Keble Road, Oxford, OX1 3RH, UK\\
$^{3}$Blackett Laboratory, Imperial College London, London SW7 2AZ, UK\\
$^{4}$School of Physics and Astronomy, University of Nottingham, Nottingham NG7 2RD\\
$^{5}$International Centre for Radio Astronomy (ICRAR), University of Western Australia, Crawley, WA6009, Australia\\
$^{6}$(SUPA) School of Physics \& Astronomy, University of St Andrews, North Haugh, St Andrews, KY16 9SS, UK\\
$^{7}$Department of Physics and Astronomy, University of Canterbury, Private Bag 4800, Christchurch, New Zealand\\
$^{8}$School of Mathematics and Physics, University of Tasmania, Private Bag 37, Hobart, TAS 7001, Australia\\
$^{9}$Jeremiah Horrocks Institute, University of Central Lancashire, Preston, PR1 2HE, UK\\
$^{10}$School of Physics \&\ Astronomy, Cardiff University, Queens Buildings, The Parade, Cardiff, CF24 3AA, UK\\
$^{11}$Sterrenkundig Observatorium, Universiteit Gent, Krijgslaan 281–S9, Gent, 9000, Belgium\\
$^{12}$Departamento de Astrof\'{\i}sica, Facultad de CC. F\'{\i}sicas, Universidad Complutense de Madrid, E-28040 Madrid, Spain\\
$^{13}$Department of Physics and Astronomy, University of California, Irvine, CA 92697, USA\\
$^{14}$INAF-Osservatorio Astronomico di Padova, Vicolo Osservatorio 5, I-35122 Padova, Italy\\
$^{15}$SISSA, Via Bonomea 265, I-34136 Trieste, Italy\\
$^{16}$UK Astronomy Technology Centre, Royal Observatory, Edinburgh, EH9
3HJ, UK\vspace{-0.2in}}

\begin{document}

\maketitle

\def \aj {AJ}
\def \mnras {MNRAS}
\def \pasp {PASP}
\def \apj {ApJ}
\def \apjs {ApJS}
\def \apjl {ApJL}
\def \aap {A\&A}
\def \nat {Nature}
\def \araa {ARAA}
\def \iaucirc {IAUC}
\def \aaps {A\&A Suppl.}
\def \qjras {QJRAS}
\def \na {New Astronomy}
\def \aapr {A\&ARv}
\def\lesssim{\mathrel{\hbox{\rlap{\hbox{\lower4pt\hbox{$\approx$}}}\hbox{$<$}}}}
\def\gtrsim{\mathrel{\hbox{\rlap{\hbox{\lower4pt\hbox{$\approx$}}}\hbox{$>$}}}}


\begin{abstract}
We use multi-wavelength (0.12 - 500 $\mu$m) photometry from
\emph{Herschel}-ATLAS, WISE, UKIDSS, SDSS and GALEX, to study 23
nearby spheroidal galaxies with prominent dust lanes (DLSGs).
DLSGs are considered to be remnants of recent minor mergers,
making them ideal laboratories for studying both the interstellar
medium (ISM) of spheroids and minor-merger-driven star formation
in the nearby Universe. The DLSGs exhibit star formation rates
(SFRs) between 0.01 and 10 M$_{\odot}$ yr$^{-1}$, with a median of
0.26 M$_{\odot}$ yr$^{-1}$ (a factor of 3.5 greater than the
average SG). The median dust mass, dust-to-stellar mass ratio and
dust temperature in these galaxies are around 10$^{7.6}$
M$_{\odot}$, $\approx$0.05\% and $\approx$19.5 K respectively. The
dust masses are at least a factor of 50 greater than that expected
from stellar mass loss and, like the SFRs, show no correlation
with galaxy luminosity, suggesting that both the ISM and the star
formation have external drivers. Adopting literature gas-to-dust
ratios and star formation histories derived from fits to the
panchromatic photometry, we estimate that the median current and
initial gas-to-stellar mass ratios in these systems are
$\approx$4\% and $\approx$7\% respectively. If, as indicated by
recent work, minor mergers that drive star formation in spheroids
with $(NUV-r)>3.8$ (the colour range of our DLSGs) have stellar
mass ratios between 1:6 and 1:10, then the satellite gas fractions
are likely $\geq$50\%.
\end{abstract}


\begin{keywords}
galaxies: formation -- galaxies: evolution -- galaxies: interactions --
galaxies: ISM -- galaxies: elliptical and lenticular, cD \vspace{-0.2in}
\end{keywords}


\section{Introduction}
Spheroidal galaxies (SGs) dominate the massive-galaxy census in
the local Universe making them good laboratories for studying
galaxy evolution at late epochs. While largely composed of old
stars \citep[e.g.][]{Trager2000a}, recent work on SGs using
ultraviolet (UV) photometry demonstrates widespread recent star
formation \citep[e.g.][]{Kaviraj2007a}, which builds $\approx$20\%
of their stellar mass after $z \approx1$ \citep{Kaviraj2008}. A
strong correspondence between blue UV colours and morphological
disturbances suggests that the star formation is merger-driven.
However, the major-merger rate is far too low to satisfy the
frequency of morphologically disturbed SGs, indicating that
\emph{minor} mergers (mass ratios $<$ 1:3) with gas-rich dwarfs
\citep{Kaviraj2009c,Kaviraj2011} drives the star formation in
these systems. SGs with prominent dust lanes (DLSGs) are a rare
class of object, considered to be the remnants of \emph{recent}
minor mergers (Shabala et al. 2012, see also Hawarden et al. 1991;
Oosterloo et al. 2002; Michel-Dansac et al. 2008; Alonso et al.
2010). These systems are good laboratories for studying both the
inter-stellar medium (ISM) in SGs and the minor-merger process
that dominates star formation in SGs at late-epochs.

While a detailed study of DLSGs is a compelling project, given
their rarity, large samples of these galaxies have traditionally
not been available. However, \citet[][K12 hereafter]{Kaviraj2012}
have recently performed a survey-scale study of DLSGs, drawn from
the entire Sloan Digital Sky Survey (SDSS) Data Release 7
\citep{Abazajian2009} via the Galaxy Zoo (GZ) project
\citep{Lintott2008}. {\color{black} The K12 dust lanes are
extended features (comparable to the galaxy effective radii) and
ideally require colour images for detection, since the lane alters
the ambient colour more dramatically than the brightness.} K12
have demonstrated that, compared to the average SG, DLSGs reside
in lower density environments, exhibit bluer UV-optical colours,
an order of magnitude higher incidence of AGN and disturbed
morphologies even in the shallow (54s exposure) SDSS images,
indicating a recent merger. Neither the dust masses in the
prominent lanes nor the `warm' dust masses (calculated via
\emph{Infrared Astronomical Satellite} [IRAS] fluxes), show any
correlation with galaxy luminosity or the age of the recent
starburst, indicating that the dust has an external origin.

Notwithstanding the detailed treatment in K12, an accurate
analysis of star formation and the ISM in DLSGs greatly benefits
from a study involving \emph{Herschel} \citep{Pilbratt2010}. For
example, star formation rates (SFRs) in weakly star-forming
systems like SGs cannot be reliably measured using SDSS H$\alpha$
fluxes because the dominant sources of ionisation are often AGN or
evolved stars \citep[e.g.][]{sauron5}, and also because the 3"
spectroscopic fibre samples varying fractions of light in
individual galaxies (corrections can be made for this, but are
inherently uncertain). Similarly, since \emph{IRAS} largely traces
warm ($>$25 K) dust \citep[e.g.][]{Dunne2001}, previous studies
based on IRAS data may not reflect the true ISM characteristics of
nearby SGs. In this Letter, we combine far-infrared (FIR)
photometry from \emph{Herschel} with UV (\emph{Galaxy Evolution
Explorer} [GALEX]; Martin et al. 2005; Gil de Paz et al. 2007),
optical (SDSS; Abazajian et al. 2009), near-infrared (\emph{UKIRT
Infrared Deep Sky Survey} [UKIDSS]; Lawrence et al. 2007) and
mid-infrared (\emph{Wide Field Infrared Sky Explorer} [WISE];
Wright et al. 2010) data, to perform a panchromatic study of DLSGs
with an unprecedentedly wide wavelength baseline (0.12 - 500
$\mu$m). The aims are to (a) study the physical properties of
DLSGs (e.g. SFRs, dust masses, dust fractions) and (b) reconstruct
the gas properties of the progenitor systems to improve our
understanding of the minor-merger process.

This Letter is organised as follows. In \S 2, we briefly describe the
original K12 sample, the \emph{Herschel}-overlap subset of K12 galaxies that
underpins this study and the methodology for deriving the physical
properties of DLSGs using panchromatic photometric data. In \S 3, we present
these derived properties and in \S 4 we reconstruct the gas properties of
the progenitor systems. We summarise our findings in \S 5. Throughout, we
use the \citet{Komatsu2011} cosmological parameters and present photometry
in the AB magnitude system \citep{Oke1983}.

\begin{figure}
\begin{center}
\includegraphics[width=2in]{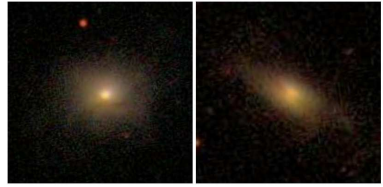}
\caption{SDSS $gri$ composite images of typical DLSGs in our study. Note the
widespread morphological disturbances that are apparent even in the shallow
(54s exposure) SDSS images.} \label{fig:all_images} \vspace{-0.2in}
\end{center}
\end{figure}

\begin{figure}
\begin{center}
$\begin{array}{c}
\includegraphics[width=2.5in]{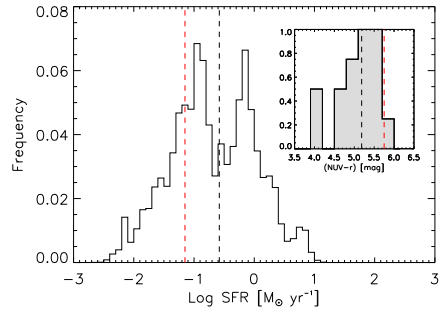}
\end{array}$
\caption{MAIN PANEL: DLSG star formation rates, estimated using
panchromatic (0.12 - 500 $\mu$m) photometry. The black and red
dashed lines indicate median values for the DLSGs and the general
SG population (Rowlands et al. 2012) respectively. INSET: The
$(NUV-r)$ colours of the DLSGs. The black and red dashed lines
indicate median values for the DLSGs and the general SG population
(Kaviraj et al. 2007) respectively.} \label{fig:sfr}
\vspace{-0.2in}
\end{center}
\end{figure}


\vspace{-0.2in}
\section{Galaxy sample and parameter estimation}
The K12 sample is drawn from the entire SDSS Data Release 7 (which
covers 11663 deg$^{2}$) through visual inspection of galaxy images
via the Galaxy Zoo (GZ) project. GZ is a unique tool that has used
300,000+ members of the public to morphologically classify, via
visual inspection, the entire SDSS spectroscopic sample
($\approx$1 million objects). GZ is uniquely powerful in its
ability to identify \emph{rare} objects such as mergers
\citep{Darg2010} and dust lanes (e.g. K12) that are best selected
via visual inspection, and has produced large samples of such
objects that are ideal for followup using other facilities.

{\color{black}Briefly, ~19,000 GZ galaxies at $z < 0.1$ were
flagged as containing a dust lane by \emph{at least} one GZ user
(individual galaxies have 50+ classifications). Each galaxy in
this sample was visually re-inspected by SK and YST to determine
whether the galaxy really has a dust feature and to select systems
with early-type morphology. This yields a final sample of 352
DLSGs, with a median $z$, absolute $r$-band magnitude and stellar
mass of 0.04, -21.5 and 10$^{10.8}$ M$_{\odot}$ respectively. To
check completeness, 1000 random SDSS SGs were inspected by SK and
YST to determine how many DLSGs were \emph{not} identified as such
by at least one GZ user. This test yielded no results. In other
words \emph{all} DLSGs are classified as such by at least one GZ
user and the K12 sample is likely complete for systems with
prominent dust lanes.}

\begin{figure*}
\begin{minipage}{172mm}
\begin{center}
$\begin{array}{cc}
\includegraphics[width=2.4in]{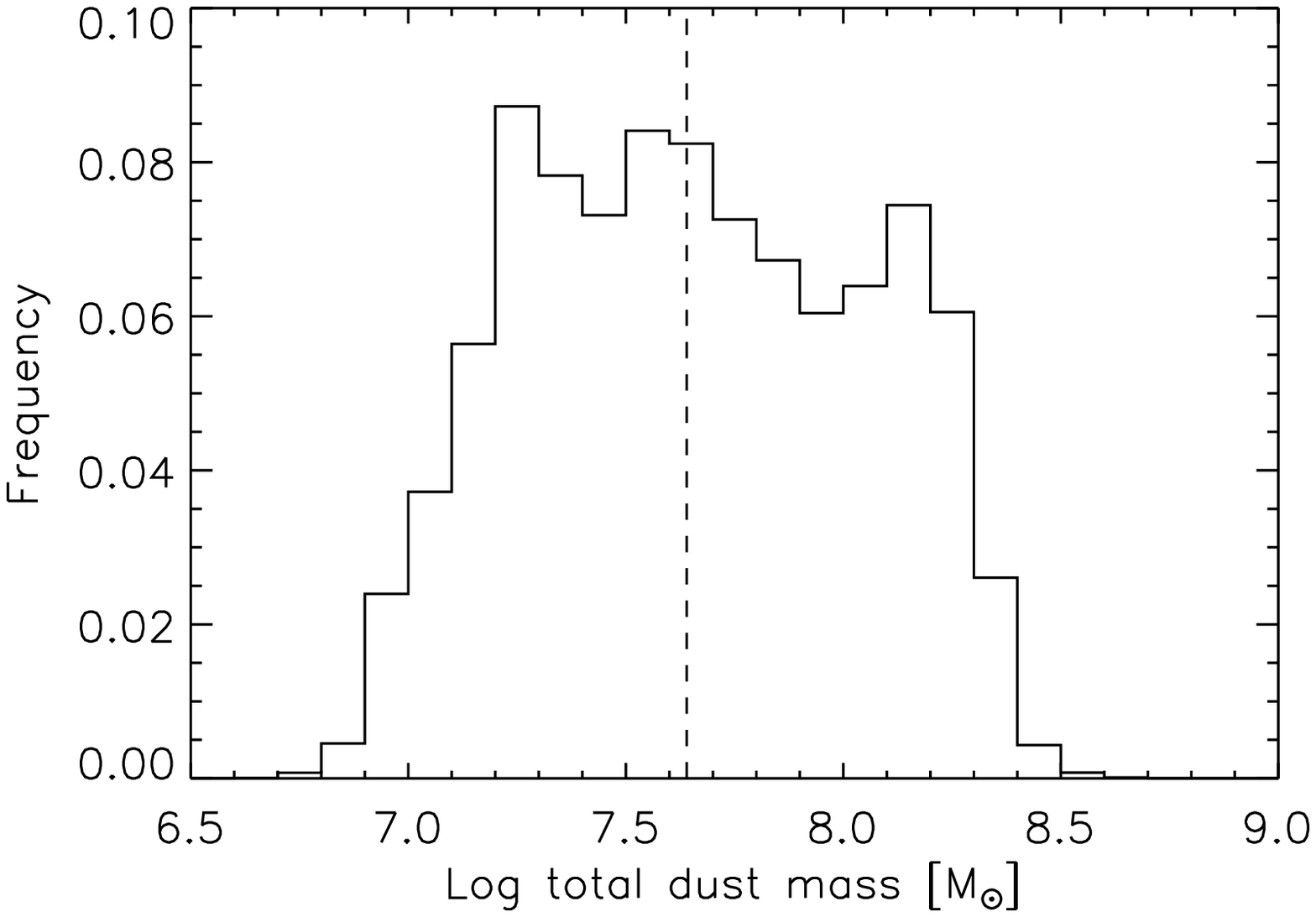} & \includegraphics[width=2.4in]{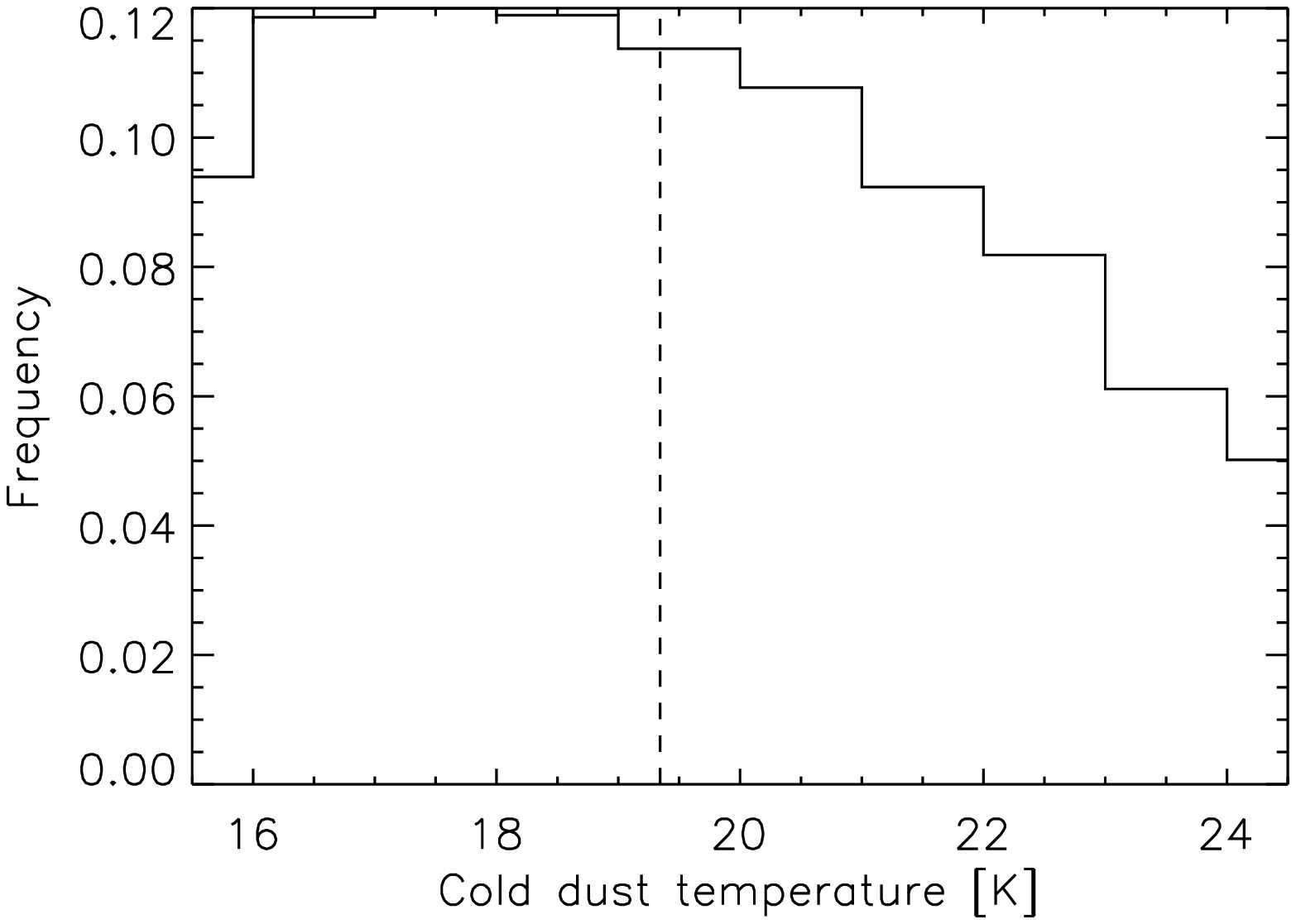}\\
\includegraphics[width=2.4in]{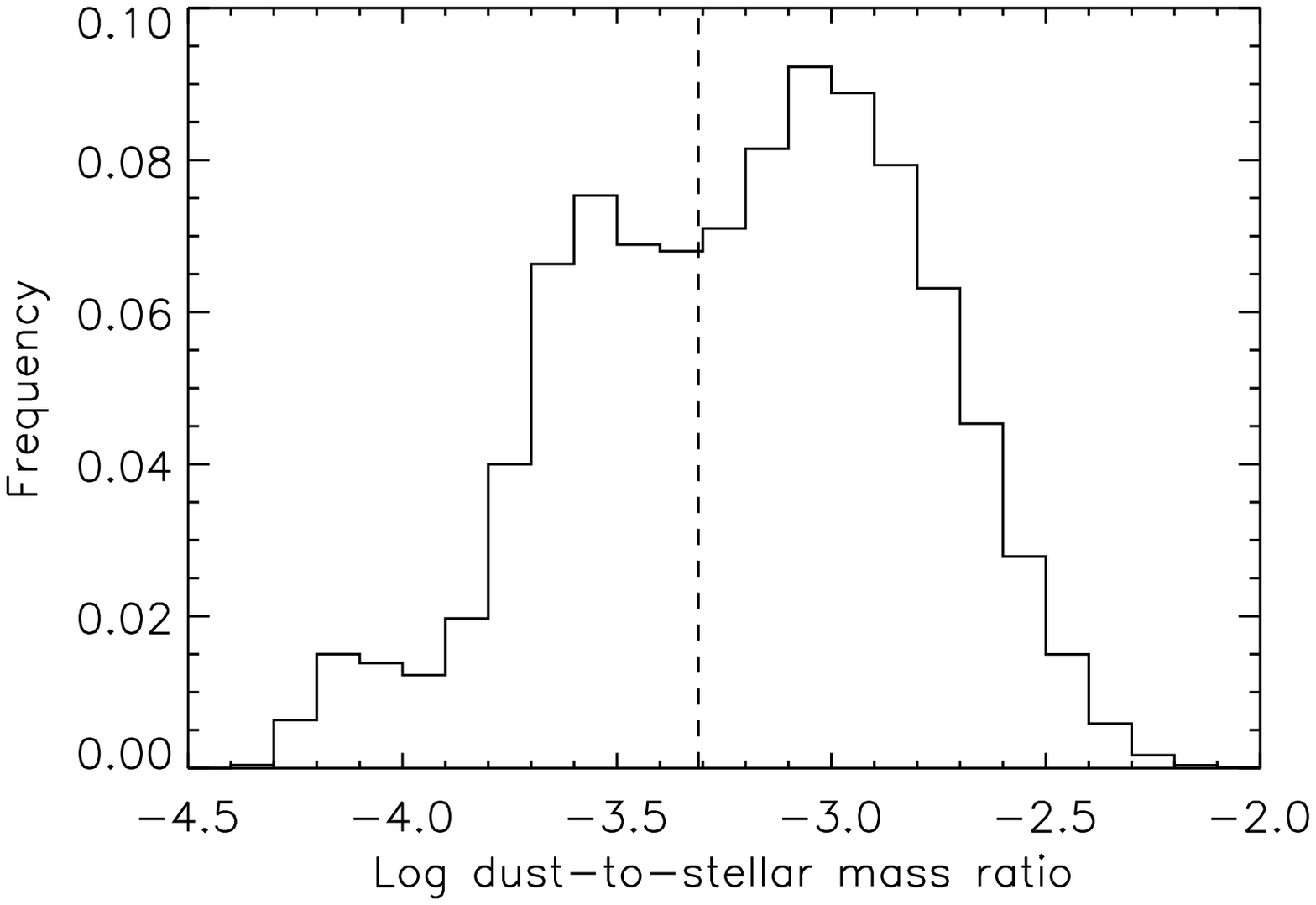} & \includegraphics[width=2.4in]{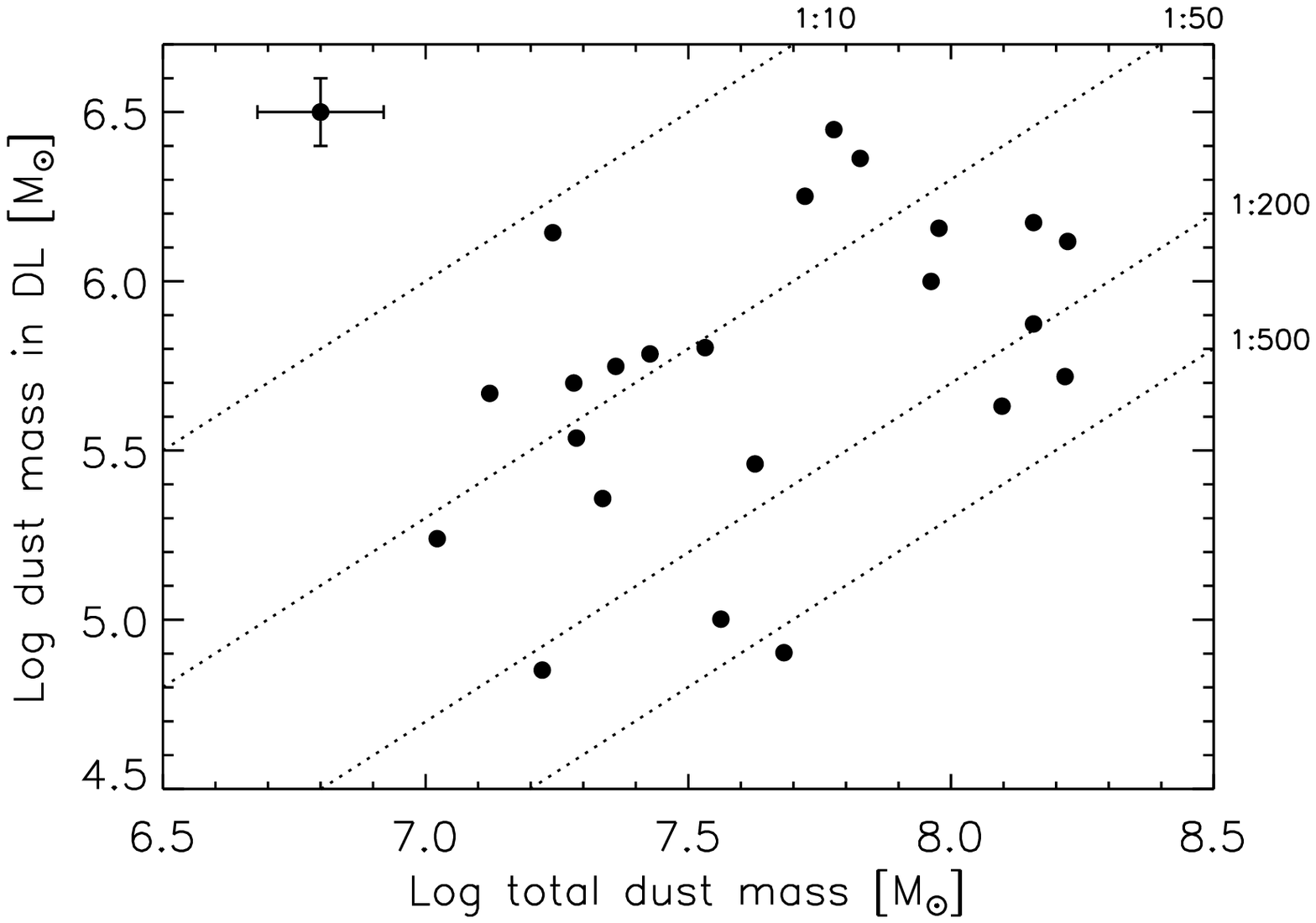}
\end{array}$
\caption{Total dust mass (top left), cold dust temperature (top
right), dust-to-stellar mass ratio (bottom left) and dust mass
hosted by the dust lanes vs total dust mass (bottom right). Dashed
lines indicate median values. The dotted lines in the bottom-right
panel indicate constant mass ratios. The histograms are stacked
PDFs for the DLSG population as a whole (see text for details).}
\label{fig:dust_properties1} \vspace{-0.2in}
\end{center}
\end{minipage}
\end{figure*}

The DLSGs in this study are extracted by cross-matching the K12
objects with the Phase 1 and North Galactic Pole (NGP) fields of
the \emph{Herschel}-ATLAS survey \citep[H-ATLAS;][]{Eales2010}, a
550 deg$^{2}$ survey at 100 - 500 $\mu$m using \emph{Herschel's}
SPIRE \citep{Griffin2010} and PACS \citep{Poglitsch2010}
instruments. We refer readers to \citet{Pascale2011} and
\citet{Ibar2010} for details of the map making. These fields cover
$\approx$50\% of the full H-ATLAS footprint, and all DLSGs in
these fields are detected at $>5\sigma$ at 250$\mu$m
\citep{Rigby2011}, with reliable optical counterparts to the
sub-mm emission (Smith et al. 2011; Hoyos et al. in prep; Valiante
et al. in prep). In addition, we leverage panchromatic data from
several overlapping surveys. From UV to $K$-band, we exploit the
`Galaxy and Mass Assembly' project
\citep{Robotham2010,Baldry2010,Driver2011}, that provides $r$-band
matched aperture photometry from GALEX (UV), SDSS (optical) and
UKIDSS (near-infrared) \citep{Hill2011}. In the mid-infrared (3.4
- 22 $\mu$m), we exploit the WISE All-Sky Survey
\citep{Wright2010}. The final DLSG sample comprises 23 galaxies,
with a median redshift and stellar mass of 0.05 and 10$^{10.9}$
M$_{\odot}$ respectively, and 20-filter photometry covering three
orders of magnitude in wavelength (0.12 - 500 $\mu$m). Figure
\ref{fig:all_images} presents SDSS images of typical galaxies in
our sample.

We estimate the physical properties of our DLSGs using the
energy-balance technique of da Cunha, Charlot and Elbaz (2008;
DC08 hereafter) via the \texttt{MAGPHYS} code. Briefly, DC08
compute a library of model spectra, described by dust emission
from stellar birth clouds (with contributions from polyaromatic
hydrocarbons (PAHs) and hot [130-250 K], warm [30-60 K] and cold
[15-25 K] grains), and match it to a second model library,
generated using a wide range of stochastic star formation
histories, metallicities and dust attenuations. The matching is
performed by assuming that the energy absorbed by dust (described
by the second library) is re-emitted in the infrared wavelengths
(described by the first library). Stellar populations are modelled
using the latest version of the \citet{BC2003} stellar models,
with dust attenuation calculated using \citet{Charlot2000}. We
refer readers to DC08 for full details of the method. The
properties of individual galaxies are estimated by comparing their
observed multi-wavelength photometry to the matched DC08 model
spectra and extracting marginalised probability density functions
(PDFs) for parameters of interest -- e.g. stellar masses, SFRs
(averaged over the last 10$^7$ yr) and dust masses -- from which
median likelihood estimates can be calculated. Note that the
20-band (UV to far-infrared) wavelength baseline employed here
significantly improves the accuracy of the derived parameters
compared to all previous studies of DLSGs.


\vspace{-0.2in}
\section{Physical properties of spheroids with dust lanes}
We begin by discussing the physical properties of our DLSGs.
Figure \ref{fig:sfr} (inset) indicates that the median $(NUV-r)$
colour of these systems is $\approx$0.5 mag bluer than the general
SG population (in the same redshift and mass range), due to star
formation. The SFRs in these systems (Figure \ref{fig:sfr}, main
panel) span the range 0.01 - 10 M$_{\odot}$ yr$^{-1}$, with a
median value of $\approx$0.26 M$_{\odot}$ yr$^{-1}$, which is a
factor of 3.5 greater than that in the general SG population
\citep{Rowlands2012}. Figure \ref{fig:dust_properties1} presents
the derived dust properties of our DLSGs. We show stacked PDFs for
the DLSG population as a whole, calculated by summing the
marginalised parameter PDFs for individual galaxies and
normalising by the total number of DLSGs. The median dust mass in
these systems is around 10$^{7.6}$ M$_{\odot}$, an order of
magnitude higher than the typical (\emph{Herschel}-derived) dust
mass in the general SG population
\citep[see][]{Rowlands2012,Smith2012}. The median dust temperature
is $\approx$19.5 K and the typical dust-to-stellar mass ratio is
$\approx$0.05\%.

In the bottom-right panel of Figure \ref{fig:dust_properties1}, we show the
ratio of the dust mass in the prominent dust lanes (calculated by K12) to
the total dust mass (measured using our multi-wavelength analysis). Briefly,
the dust mass in the lanes is calculated as follows. For each galaxy,
ellipses are fitted to isophotes in the SDSS $r$-band image (with the dust
features masked out), using the \texttt{IRAF} task \texttt{ELLIPSE}. For
each galaxy, an extinction map is constructed using the ratio of the
observed flux and the flux in the model fit (calculated by fitting ellipses
to the SDSS $r$-band image with the dust features masked out). The Galactic
mass absorption coefficient is then used to calculate the dust mass that
resides in the large-scale features. We refer readers to Section 6.1 of K12
for details of this method. Figure \ref{fig:dust_properties1} indicates that
the dust fraction hosted by the prominent dust lanes in these systems is
remarkably small, between 1/10 and 1/500 of the total dust mass (with a
median value of $\approx$1/50 i.e. $\approx$2\%).

\begin{figure}
$\begin{array}{c}
\includegraphics[width=2.4in]{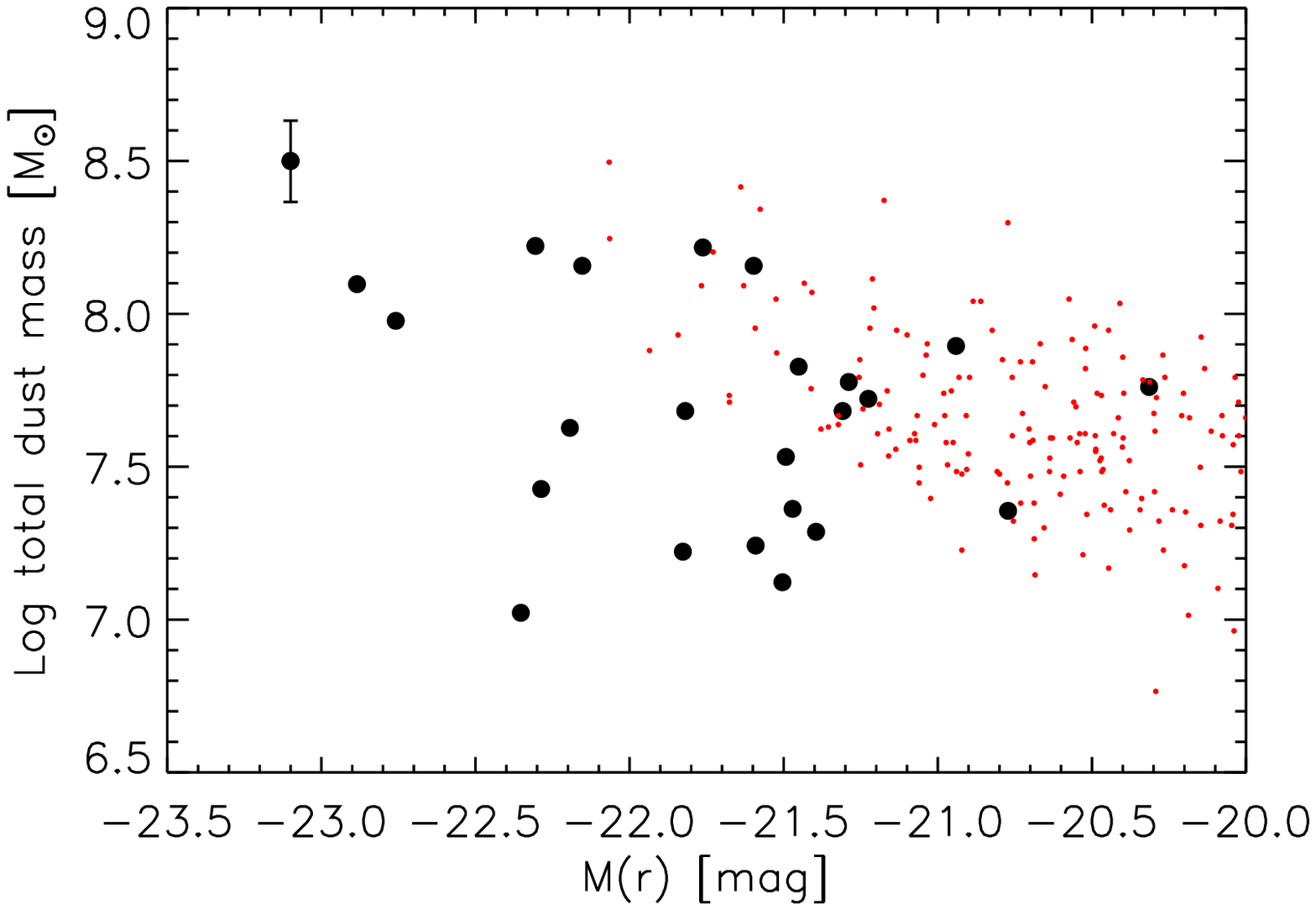}\\
\includegraphics[width=2.4in]{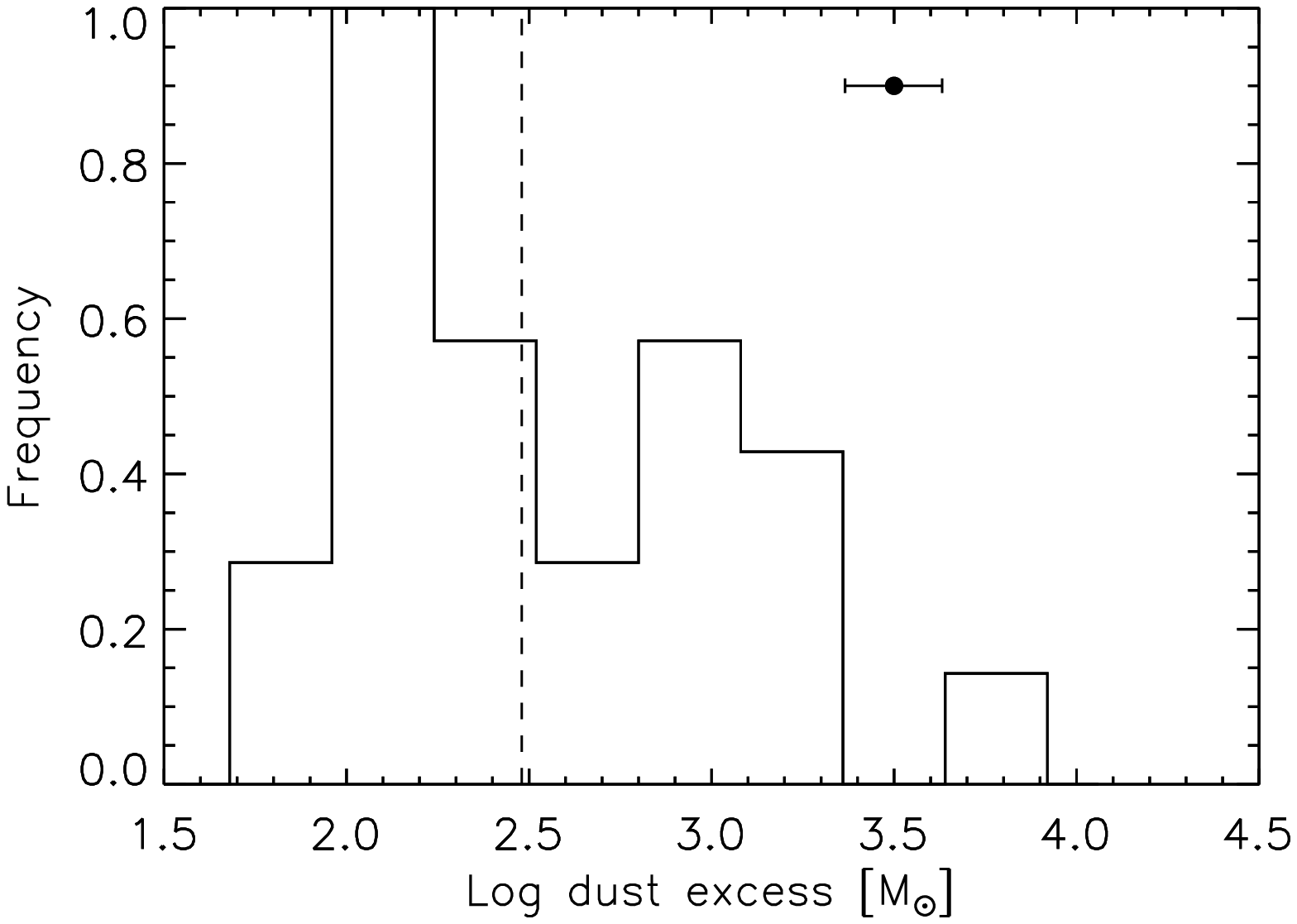}
\end{array}$
\caption{TOP: Total dust mass vs. M(r). Black circles are DLSGs
and small red circles indicate spirals. BOTTOM: Distribution of
dust excess in our DLSGs, defined as the difference between the
total dust mass and that expected from stellar mass loss. The
median value is indicated by the vertical dashed line.}
\label{fig:dust_properties2} \vspace{-0.1in}
\end{figure}

We explore the origin of the dust in our DLSGs in Figure
\ref{fig:dust_properties2}. The top panel shows that there is no
correlation between the total dust masses and absolute $r$-band
magnitudes of our galaxies. By contrast, the dust masses in spiral
galaxies correlate well with their luminosities (indicating an
internal origin). Our results remain the same regardless of the
optical filter used. Note that the DLSG SFRs also show a similar
lack of correlation with absolute magnitude (not shown). In the
bottom panel of Figure \ref{fig:dust_properties2}, we plot the
distribution of the `dust excess' in individual DLSGs, defined as
the difference between the total dust mass and the maximum dust
mass expected from internal stellar mass loss. The mass-loss
contribution is estimated using the blue luminosity following
(Merluzzi et al. 1998; see their Figure 1) and assumes dust
depletion via sputtering in the hot gas over a maximal destruction
time-scale of 10$^{7.5}$ yr \citep[see
e.g.][]{Draine1979,Dwek1990}, although it is worth noting here
that empirical dust depletion rates have not been extensively
studied in SGs. The dust masses in our DLSGs are at least a factor
of 50 greater than the dust mass expected from stellar mass loss.
Taken together, these results argue against an internal origin for
the dust content and indicate that both the ISM and star formation
in these galaxies have \emph{external} drivers. The widespread
morphological disturbances further suggest that this driver is
indeed a recent merger. {\color{black}It is worth noting here that
DLSGs inhabit low-density environments (only 2\% of the K12 sample
are in clusters). The Herschel-detected DLSGs in this paper
exclusively inhabit the field, so that environmental processes
like ram-pressure stripping of the ISM are likely to be absent in
these galaxies.}


\vspace{-0.2in}
\section{Gas properties of the progenitor (minor) mergers}
Given the significant role of minor mergers in driving star formation in
massive galaxies at late epochs \citep[e.g.][]{Kaviraj2011}, empirical
constraints on the gas properties of these minor mergers are desirable but
remain scant. Our DLSGs offer an opportunity to explore these properties, in
particular the gas fractions of the satellites that fuel star formation in
massive galaxies at late epochs. While this was explored in K12, the
significantly higher quality of both the data and the derived dust
properties in this study allows us to reconstruct the gas fraction of the
progenitor systems of our DLSGs with better accuracy.

We begin by estimating the \emph{current} gas-to-stellar mass ratios in our
DLSGs, using recently measured gas-to-dust (G/D) ratios in SGs, that involve
\emph{Herschel}-derived dust masses. Note that `gas' refers to the sum of
the atomic (\texttt{HI}) and molecular ($H_2$) gas masses. In Cen A -- the
archetypal SG with a dust lane, and a system that is similar to our DLSGs --
\citet{Parkin2012} calculate a G/D ratio of 103$\pm$8. The equivalent values
in nearby SGs from the \emph{Herschel} Reference Survey (which do not have
prominent dust lanes) lie in the range 120$\pm$24 \citep{Smith2012}. The
values measured so far thus appear to lie within a reasonably small range
($\approx$100 to 150), and close to the canonical Milky Way value of 150
\citep[][]{Spitzer1978,Draine1984}, regardless of whether the SG in question
hosts a prominent dust lane. This appears consistent with the fact that only
a small fraction of the total dust mass is hosted by the dust lanes
themselves. Using a mean G/D ratio of 120 then yields \emph{current}
gas-to-stellar mass ratios in our DLSGs between 1 and 20\%, with a median of
$\approx$4\%. As star formation is clearly ongoing in these systems, these
gas-to-stellar mass ratios are lower than the \emph{initial} ratios in the
progenitor systems.

Since the native cold gas in SGs is negligible and incapable of driving the
observed UV colours of our DLSGs \citep[see e.g.][]{Kaviraj2007a}, and given
the evidence for the ISM having an external origin, it is reasonable to
assume that the gas in the progenitor system is provided by the accreted
satellite. We conclude this section by estimating the initial gas-to-stellar
mass ratios of the DLSGs and the gas fractions of satellites that are
fuelling their star formation.

\begin{figure}
\begin{center}
\includegraphics[width=2.4in]{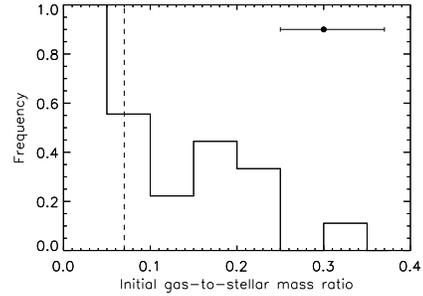}
\caption{Initial gas-to-stellar mass ratios in the progenitor
systems of our DLSGs (see text in Section 4 for details). The
median value is indicated by the vertical dashed line.}
\label{fig:gas}\vspace{-0.1in}
\end{center}
\end{figure}

Detailed numerical simulations \citep[see][]{Kaviraj2009c}
indicate that SGs with $NUV-r>3.8$ (the colour range spanned by
our DLSGs) are likely remnants of mergers with mass ratios between
1:6 and 1:10, and that the induced star formation is typically
complete within $\approx$3 Gyr (largely independent of mass
ratio). Thus, the initial gas-to-stellar mass ratio can be
estimated by integrating the \texttt{MAGPHYS}-derived SFH of each
DLSG over the last 3 Gyr, adding this to the current gas mass and
dividing by the stellar mass of the system. Figure \ref{fig:gas}
presents these initial gas-to-stellar mass ratios, with the dashed
line indicating the median value ($\approx$7\%). Finally, given
mass ratios between 1:6 and 1:10, a median gas-to-stellar mass
ratio of $\approx$7\% implies that the gas fractions in the
satellites accreted by our DLSGs were likely to be high, between
50 and 80\%.\footnote{If is $g$ is the system gas mass and $m$ and
$M$ are the stellar masses of the satellite and spheroid, then the
initial gas-to-stellar mass ratio ($k$) is: $k=\frac{g}{m+M}$.
Assuming the satellite donates the gas then the satellite gas
fraction is $\frac{g}{m}=k\times(1+\frac{M}{m})$.}. Note that,
given the assumptions employed in deriving these estimates, the
values are only indicative - a more accurate analysis of the gas
properties can be achieved using longer wavelength follow up of
the DLSGs, using facilities such as the IRAM radio telescope and
the Atacama Large Millimetre Array (ALMA).


\vspace{-0.2in}
\section{Summary}
We have studied 23 local ($\langle z\rangle\approx$ 0.05), massive ($\langle
\textnormal{M}_*\rangle\approx 10^{10.9} \textnormal{M}_{\odot}$) spheroidal
galaxies with prominent dust lanes (DLSGs), drawn from the SDSS DR7 via
visual inspection using the Galaxy Zoo project. The DLSGs studied here have
20-band (0.12 - 500 $\mu$m) photometric data from the \emph{Herschel}-ATLAS
(far-infrared), WISE (mid-infrared), UKIDSS (near-infrared), SDSS (optical)
and GALEX (UV) surveys, allowing us to probe the properties of their
inter-stellar media (ISM) and star formation in unprecedented detail. DLSGs
are known to be the remnants of recent \emph{minor} mergers between
spheroids and gas-rich dwarfs - given the important contribution of minor
mergers in driving cosmic star formation at late epochs, these systems are,
therefore, unique laboratories for studying both the ISM of spheroids and
the minor-merger process in the local Universe.

Our DLSGs are $\approx$0.5 mag bluer in the $(NUV-r)$ colour than
the general SG population, due to enhanced star formation. The
star formation rates span a large range (0.01 - 10 M$_{\odot}$
yr$^{-1}$), with a median value of 0.26 M$_{\odot}$ yr$^{-1}$. The
median dust mass is $\approx$10$^{7.6}$ M$_{\odot}$, an order of
magnitude greater than the typical dust mass in the average SG.
The median dust-to-stellar mass ratio is $\approx$0.05\% and the
typical dust temperature is $\approx$19.5 K. The dust hosted by
the prominent dust lanes is remarkably small, between 1/10 and
1/500 (with a median of 1/50) of the total dust mass in the
system. The dust masses and star formation rates show no
correlation with galaxy luminosity, and the dust masses are at
least a factor of 50 greater than the contribution expected from
stellar mass loss. Taken together, this indicates an external
origin and driver for the ISM and star formation in these systems.

The DLSGs offer a route for estimating the gas properties of the minor
mergers that drive star formation in massive galaxies in the nearby
Universe. Using literature gas-to-dust ratios, the \emph{current}
gas-to-stellar mass ratios in these systems are estimated to be between 1
and 20\%, with a median value of 4\%. Combining these current ratios with
star formation histories derived using panchromatic (0.12 - 500 $\mu$m)
photometry, we have estimated the \emph{initial} gas-to-stellar mass ratios
in the progenitor systems of our DLSGs to be $<$30\%, with a median value of
7\%. If -- as suggested by recent work -- the mass ratios of minor mergers
that induce star formation in SGs with $(NUV-r)>3.8$ is between 1:6 and
1:10, then the gas fractions of the satellites accreted by our DLSGs were
likely to be $>$50\%.

In forthcoming work, we will further probe the properties and external
origin of the ISM in these systems e.g. by using IRAM to derive accurate G/D
ratios which correlate with the gas-phase metallicity and using
integral-field spectroscopy to trace kinematical misalignments between the
gas and stars, which would be expected if the gas is accreted from an
external source.


\vspace{-0.2in}
\section*{Acknowledgements}
We thank the anonymous referee for constructive comments that
improved the paper and Sperello di Serego Alighieri, Noah Brosch
and Marc Sarzi for interesting discussions. S.K. acknowledges
fellowships from Imperial College London and Worcester College
Oxford. The Herschel-ATLAS is a project with \emph{Herschel},
which is an ESA space observatory with science instruments
provided by European-led Principal Investigator consortia and with
important participation from NASA. The H-ATLAS website is
http://www.h-atlas.org/. GAMA is a joint European-Australasian
project based around a spectroscopic campaign using the
Anglo-Australian Telescope. The GAMA input catalogue is based on
data taken from the Sloan Digital Sky Survey and the UKIRT
Infrared Deep Sky Survey. Complementary imaging of the GAMA
regions is being obtained by a number of independent survey
programs including \emph{GALEX} MIS, VST KIDS, VISTA VIKING,
\emph{WISE}, \emph{Herschel}-ATLAS, GMRT and ASKAP providing UV to
radio coverage. GAMA is funded by the STFC (UK), the ARC
(Australia), the AAO, and the participating institutions. The GAMA
website is http://www.gama-survey.org/.


\vspace{-0.2in}
\nocite{dacunha2008,Merluzzi1998,Martin2005,Abazajian2009,Lawrence2007,Wright2010,Smith2011,GildePaz2007,
Oosterloo2002, Alonso2010, Michel-Dansac2008} \small
\bibliographystyle{mn2e}
\bibliography{references}


\end{document}